\documentclass[11pt,letterpaper]{article}

\title{On the Applicability of Combinatorial Designs to\\Key Predistribution for Wireless Sensor Networks}

\author{Keith~M.~Martin\\
Information Security Group, Royal Holloway, University of London,\\
Egham, Surrey TW20 0EX, U.K.}

\newtheorem{definition}{Definition}[section]

\newcommand{\B}{\cal{B}}

\newcommand{\E}{\cal{E}}
\newcommand{\G}{\cal{G}}
\newcommand{\HP}{\cal{H}}
\newcommand{\I}{\cal{I}}
\newcommand{\U}{\cal{U}}
\newcommand{\eS}{\cal{S}}

\begin{document}

\maketitle

\begin{abstract}
The constraints of lightweight distributed computing environments such as wireless sensor networks lend themselves to the use of symmetric cryptography to provide security services. The lack of central infrastructure after deployment of such networks requires the necessary symmetric keys to be predistributed to participating nodes. The rich mathematical structure of combinatorial designs has resulted in the proposal of several key predistribution schemes for wireless sensor networks based on designs. We review and examine the appropriateness of combinatorial designs as a tool for building key predistribution schemes suitable for such environments.
\end{abstract}

\section{Introduction}

 The management of cryptographic keys in any information system is one of the most challenging aspects of implementing cryptography. One of the most important key management processes is \textit{key establishment}, which governs the placement of cryptographic keys in a network. This is especially relevant in applications of symmetric cryptography, where it is necessary to ensure that all parties who are authorised to access (or verify) a cryptographically protected piece of information have the appropriate key.

 Symmetric key establishment almost always involves a trusted third party, which we will term a \textit{key management authority} (KMA), at some stage in the process. In some environments this KMA is online and available at time of use. In such cases the third party is often referred to as a \textit{key distribution centre}. However in many other environments it is not possible for a KMA to form part of a live network and assist in online key establishment. In this case the KMA can only be involved in initialisation processes that take place prior to deployment of the network. At this stage the KMA must equip each node in the network with the necessary cryptographic keys for facilitating security services after the nodes are deployed in the network. Key establishment schemes of this type are usually referred to as \textit{key predistribution schemes} (KPSs) because the keys are distributed in advance and cannot be generated ``on the fly''.

 A major current trend in computing technologies is a shift from centralised, relatively stable, wired networks consisting of powerful devices, to distributed, dynamic (ad hoc), wireless networks consisting of lightweight devices. This is being driven by the development of very small wireless computers, which can either be deployed on their own or embedded into everyday objects. The resulting ubiquitous networks have several important characteristics that typically include the need to conduct basic network services such as routing using the network nodes themselves (rather than via a centralised infrastructure), high unavailability rates of nodes, and the need for highly efficient network protocols due to the power and energy constraints of the nodes. From the perspective of providing security services, these characteristics lend themselves to the use of symmetric cryptography and to key predistribution for key establishment.

 Wireless sensor networks are just one class of emerging technologies of this type. While we will frame our discussion around wireless sensor networks, which is the context for almost all of the related research, it is worth noting that many of the schemes we discuss may be equally applicable to other technologies with similar characteristics to wireless sensor networks.

 Combinatorial structures are natural objects on which to model many aspects of symmetric key management. For a survey of their contributions to key establishment, see \cite{Martin:BCC}. In this paper we will focus only on key predistribution, and on the application of combinatorial designs in particular.

 The paper is organised as follows. In Sect.~\ref{sec:wsn} we discuss wireless sensor networks, outlining aspects which are of relevance to key predistribution. In Sect.~\ref{sec:des} we provide a brief background to combinatorial designs. In Sect.~\ref{sec:kps} we outline a basic model for a KPS and discuss fundamental schemes. In the remaining sections we look at different applications of designs to key predistribution. In Sect.~\ref{sec:direct} we discuss direct application of designs as KPSs. In Sect.~\ref{sec:building} we look at the use of designs as building blocks for KPSs. Finally in Sect.~\ref{sec:special} we focus on KPSs for special networking environments. Throughout the discussion we will consider the extent to which combinatorial designs are genuinely useful for building KPSs for wireless sensor networks.

\section{Wireless Sensor Networks}\label{sec:wsn}

A \textit{wireless sensor network} (WSN) is an ad hoc network formed from a collection of low-powered sensor nodes that gather data and use wireless communication to transmit the information that they collect. The number of nodes can vary between dozens to thousands, depending on the application \cite{romer-design}. WSNs are best suited to applications where some form of environmental monitoring is required, but where the scale and hostility of the environment does not lend itself to the deployment of a few expensive monitoring devices (such as humans). Examples include seismic data gathering, remote habitat monitoring, gathering of ecological data, forestry welfare, agriculture, disaster relief operations and military intelligence gathering \cite{irrigate,nectarine,soilnet}. The typical characteristics of a WSN are:
\begin{itemize}
\item \textit{Highly constrained nodes}. The nodes are very small battery-powered devices and are highly constrained with respect to memory storage and power. They are thus limited in their computational and communication ability.
\item \textit{Lack of central control}. Once deployed, most WSNs do not feature any central control node. Thus all network functionality must be achieved through co-operation between the nodes.
\item \textit{Requirement to form a network to a sink}. In most WSNs the assumption is that the sensor nodes will take readings and then attempt to communicate this data back to a \textit{sink}, which is a more powerful device that will periodically connect to the WSN and request data. The location of this sink in the network is typically not fixed (it could, for example, be a portable laptop).
\item \textit{Hop-based communication}. Most WSNs use radio communication to connect between nodes. The constrained nature of the nodes means that in most cases the communication range of a node will be much smaller than the network diameter. Thus nodes communicate by \textit{hopping}, meaning that a node passes data to a node within range, who then passes it onto a node within its range, etc.
\item \textit{Dynamic network structure}. It is generally assumed that WSNs are highly dynamic. Nodes are often assumed to regularly ``sleep'' to conserve battery power. Nodes expire once their battery is drained. In some WSNs the nodes are mobile, although in most current applications they are static.
\item \textit{Nodes vulnerable to compromise}. The constrained nature of sensor nodes mean that strong security protection such as tamper-resistance is usually not viable. Thus it is normally assumed that sensor nodes can be fairly easily captured and that any sensitive information (such as keys) that is stored on them is likely to be exposed.
\end{itemize}
We will make three restrictions on the type of WSN that we consider for most of this paper:
\begin{enumerate}
\item \textit{Homogeneous nodes}. We will assume that all nodes have the same capabilities and constraints.
\item \textit{Communication structure}. We will assume that the main aim of any communication in the WSN is to send data from a node to the sink. We will thus not be attempting to set up fully connected subnetworks or establish group keys.
\item \textit{No mobility}. We will assume (for simplicity) that nodes are not mobile after deployment. In fact many of the solutions discussed here are also appropriate for mobile nodes.
\end{enumerate}
An important issue that affects KPS design is that WSNs vary in the extent to which the location of nodes is known prior to deployment. We will thus follow \cite{keithframework} by classifying WSNs as being either:
\begin{enumerate}
\item \textit{Uncontrolled} if the location of sensors cannot be
predicted before deployment. This is the default WSN scenario and assumes that the application environment is sufficiently hostile that nodes cannot be positioned in any controlled way. For example, they may be released from the air over a disaster site.
\item \textit{Partially controlled} if some information about the
location of sensors is known before deployment. This might be the case when sensors are strategically released from the air in batches.
\item \textit{Fully controlled} if the precise location of sensors
is known before deployment. This is likely to be the case, for example, when sensors are deployed in a grid in a vineyard to monitor ground humidity.
\end{enumerate}
We will generally assume that a WSN is uncontrolled, however we will discuss KPSs for other types of WSN in Sect.~\ref{sec:special}.

There has been some debate about the practicality of using public key cryptography to implement security services in a WSN \cite{Lopez06}. While this may indeed become more practical (and where it is, some aspects of key establishment may become easier), the case for designing solutions that only use symmetric cryptography remains strong. Symmetric cryptography is still preferred in many modern applications which are not as resource constrained as WSNs because of the efficiency gains and the unique problems posed by management of public keys. Perhaps more compellingly, it is likely that as soon as public key cryptography is practical on a given sensor node technology, even more constrained sensor technology will be being developed where it is not. In this paper we assume that a fully symmetric solution is required.

\section{Combinatorial Designs}\label{sec:des}

In this section we briefly review some definitions and notation that
we will employ later. We refer the reader to the combinatorial
literature for further details \cite{colbournbook}.

A \textit{set system} $({\I},{\B})$ consists of a set ${\I}$ of $v$
elements (\textit{points}) and a collection ${\B}$ of subsets
(\textit{blocks}) of ${\I}$. The \textit{degree} of $x\in {\I}$ is
the number of blocks of ${\B}$ containing $x$ and $({\I},{\B})$ is
\textit{regular} if all points have the same degree $r$. The
\textit{rank} $k$ of $({\I},{\B})$ is the size of the largest block
in ${\B}$ and we say that $({\I},{\B})$ is \textit{uniform} if all
blocks have size $k$.

A regular, uniform set system with $|{\I}|=v$, $|{\B}|=b$ is known as a $(v,b,r,k)$-\textit{design}. In such designs it must be the case that $bk=vr$. A $(v,b,r,k)$-design in which every $t$ points occurs on precisely $\lambda$ blocks is known as
a $t$-$(v,b,r,k,\lambda)$-\textit{design} (we often just refer to a
$t$-$(v,k,\lambda)$-\textit{design} since $b$ and $r$ can then be
uniquely derived). In a \textit{dual} design, the roles of points and blocks are interchanged. \textit{Symmetric} designs are self-dual and thus have $v=b$, $k=r$ and every $t$ blocks meeting in $\lambda$ points. A symmetric 2-$(s^2+s+1, s^2+s+1,s+1,s+1,1)$-design is known as a
\textit{projective plane}.

A set system is a \textit{group-divisible design} GD$(n^u,k)$ if
$v=nu$ and there exists a partition ${\HP}$ of ${\I}$ into $u$
\textit{groups} of size $n$ such that:
\begin{enumerate}
\item Every $H\in {\HP}$ intersects a block $B\in{\B}$ in at most
one point;
\item Every pair of points from different groups occur together in
precisely one block.
\end{enumerate}
A \textit{transversal design} TD$(k,n)$ is a GD$(n^k,k)$ (in this case every $H\in {\HP}$ intersects a block $B\in{\B}$ in precisely one point). A TD$(t,k,n)$ is a further generalisation where the second condition is applied to sets of $t$ points, rather than pairs. A ${\rm TD}(k,n)$ is {\em resolvable}
if the blocks can be partitioned into sets ${\cal B}_1, {\cal B}_2,\ldots,{\cal
B}_s$ such that each point of the design is contained in exactly one block in
each set.  These sets are known as {\em parallel classes}.

A \textit{graph} $\G=(\I,\E)$ consists of a set of of vertices $\I$ joined by \textit{edges} in $\E$, where $\E
\subseteq \I \times \I$. We say that a pair of vertices $U$ and $V$
are \textit{adjacent} if $\{U,V\}\in \E$. The \textit{degree} of a vertex $U$
is the number of vertices adjacent to $U$. A graph is
\textit{regular of degree} $r$ if all vertices have degree $r$.
A \textit{complete $t$-partite} graph is a graph whose vertices
can be partitioned into $t$ disjoint subsets such that two
vertices are adjacent if and only if they belong to distinct
subsets. An $(n,r,\lambda,\mu)$-\textit{strongly regular graph} is a regular
graph on $n$ vertices with degree $r$ such that any two distinct vertices
have $\lambda$ common neighbours if they are adjacent and $\mu$
common neighbours if they are not adjacent.

\section{Key Predistribution Schemes for WSNs}\label{sec:kps}

In this section we provide an introduction to key predistribution for WSNs.

\subsection{Key Predistribution Stages}\label{sec:stages}

The lack of any central control nodes in a WSN means that in order to equip sensor nodes with symmetric keys, a KMA will need to load keys onto nodes prior to deployment using a KPS to determine which keys are allocated to which nodes. After deployment, two nodes will be able to use a cryptographic service on a network link (such as encryption or a MAC) if they:
\begin{enumerate}
\item are in radio communication range of one another; and
\item share at least one key.
\end{enumerate}
If either of these conditions is not met then the nodes will have to seek a path of network links connecting them such that these conditions are met on each of the intermediate hops. Key establishment in a WSN can thus be regarded as consisting of the following three stages:
\begin{enumerate}
\item \textit{Key predistribution}. The KMA chooses a KPS defined on the $n$ nodes ${\U}=\{U_1,\ldots ,U_n\}$ in the network. Following \cite{lee:sac04}, this KPS can de modelled by a set system $({\I},{\B})$ (sometimes referred to as a \textit{key ring}), where ${\I}=\{x_i\,:\,1\leq i\leq v\}$ is a set of $v$ \textit{key identifiers} and ${\B}=\{B_j\,:\,1\leq j\leq n\}$ is a set of $n$ \textit{node allocations}. For each key identifier $x_i$, the KMA randomly selects a key $K_i$. The KMA then associates each node $U_j$ in the network with a node allocation $B_j$ and issues $U_j$ with the keys $L_j=\{K_i\,:\, x_i\in B_j\}$. Note that the association of $U_j$ with $B_j$ need not be a secret, however the instantiation of $B_j$ by $L_j$ must be.
\item \textit{Shared key discovery}. If two nodes within communication range of one another wish to deploy a cryptographic service, they first need to determine if they have any keys in common. The default method is to broadcast their node allocations to one another, but more efficient techniques can sometimes be found. If they have key identifiers in common then a session key can be generated from the common keys associated with these identifiers by means of a suitable key derivation function.
\item \textit{Path-key establishment}. If two nodes fail to identify common keys during shared key discovery then they need to find a secure path between one another that employs intermediate nodes which can. Obviously, the shorter this secure path the better.
\end{enumerate}

\subsection{Requirements}\label{sec:reqs}

The main challenge in designing a KPS that is suitable for this type of environment is that a balance must be sought between competing, and to an extent contradictory, requirements:
\begin{itemize}
\item \textit{Storage}. Nodes are memory constrained and thus the number of keys stored on each node should be kept as low as possible.
\item \textit{Connectivity}. A WSN is dynamic and communication is expensive, thus each node should store sufficient keys that secure paths through the network can be established when needed. There are various different measures for connectivity that could be applied in the context of WSNs. Measures of \textit{global connectivity} assess the connectivity of the entire network as a whole. If the node allocations for any two nodes have non-empty intersection then we will refer to the network as having \textit{full connectivity}. Measures of \textit{local connectivity}, which assess the ability of nodes to form secure paths with nodes in their close neighbourhood are probably most appropriate. One such, from \cite{lee:practical}, is the probability that $U_i$ and $U_j$ have at least one key in common (i.e. $B_i\cap B_j\not=\emptyset$). This notion can be generalised to measure local connectivity with respect to secure paths of two hops or more.

\item \textit{Resilience}. Nodes are vulnerable to compromise, thus keys should be distributed in such a way that the damage caused by exposure of the keys stored on a node is controlled. It is not clear what the ``right'' measure of resilience is in a WSN. One suggested measure used in \cite{lee:IEEE} is $\mbox{fail}(s)$, which is the probability that a link between two noncompromised nodes $U_i$ and $U_j$ is affected after $s$ other nodes $\eS$ are compromised at random, where a link is \textit{affected} if $B_i\cap B_j\not=\emptyset$ and $B_i\cap B_j \subseteq \cup_{U_k\in \eS}B_k$. Another measure proposed in \cite{rujroy:ispa07} evaluates the probability that compromise of $s$ nodes exposes all the keys from at least one different (not compromised) node allocation.

\item \textit{Efficiency}. There are several processes involved in key establishment for a WSN that it may be desirable to make as efficient as possible since nodes are constrained by limited battery power. These include \textit{computation}, \textit{shared key discovery} and \textit{path-key establishment}. We will be considering the first two, but note that path-key establishment generally involves consideration of routing algorithms in WSNs, which is out of scope for our discussion.

\item \textit{Network size}. Since many applications of WSNs involve large numbers of nodes, it is important that a KPS can support a large number of nodes.
\end{itemize}
The main challenge in designing KPSs is that several of these requirements tend to compete with one another. For example, increasing the maximum number of nodes that can be supported often involves increasing the storage at each node. Also, many KPSs trade off measures of connectivity against resilience.

\subsection{Baseline Schemes}\label{sec:baseline}

There are several important baseline KPSs. Although these are not all designed for WSNs, they provide benchmark schemes that can also be used to illustrate the requirements tradeoffs.

\begin{description}
\item[Single Key KPS]
This KPS consists of a single key that is
stored by each node in the network.  It provides optimal connectivity and
storage, but has very poor resilience since all communication links are affected by a single node capture.
\\
\item[Complete Pairwise Key KPS]
In this KPS, a unique key is assigned to each pair of nodes.  This scheme has optimal connectivity and optimal resilience, since compromise of one node does not affect any pair of non-compromised nodes.  However this KPS requires each node to store $n-1$ keys, which is infeasible if $n$ is large (which will be the case in many WSNs).
\\
\item[Blom's KPS\cite{Blom85,Blundo93}]
This scheme uses a symmetric bivariate polynomial over a finite field
GF$(q)$, {i.e.} a polynomial $P(x,y)\in {\rm GF} (q)[x,y]$ with the property
that $P(i,j)=P(j,i)$ for all $i,j\in {\rm GF}(q)$.  Node $U_i$ stores the univariate polynomial $f_i(y)=P(U_i,y)$.  In order to establish a common key with $U_j$, node $U_i$ computes $K_{ij}=f_i(U_j)=f_j(U_i)$.
This process enables any two nodes to share a common key.  If $P$ has degree
$w$, then each share consists of a degree $w$ univariate polynomial hence each node must store the $w+1$ coefficients of this polynomial, which requires as much space as storing $w+1$ keys.  Blom's KPS thus has optimal connectivity and reasonable storage. It also has very simple shared key discovery, with two nodes simply needing to broadcast their identities to one another. With respect to resilience, an adversary who captures $s$ nodes, where $s\leq t$, does not learn any information about keys established between non-compromised nodes. However an adversary who captures $w+1$ or more nodes can interpolate the polynomial $P$ and hence learn all
the keys.

Note that Blom's KPS does not strictly conform to the model in Sect.~\ref{sec:stages} since each user stores secret information that allows it to generate its node allocation rather than storing the separate key identifiers. Thus it reduces storage at the cost of requiring computation in the form of polynomial evaluations each time a key identifier is established.
\\
\item[Random KPS \cite{eschenauer:acm02}]
This scheme is a probabilistic KPS, with each node drawing keys uniformly
without replacement from some finite keypool $\cal K$.  The properties of this scheme depend on the number of keys drawn and
the size of $\cal K$. In the basic scheme any two nodes can communicate securely if they share at least one key. The basic scheme was further parameterised in \cite{chan:IEEE03}, where an additional threshold parameter was introduced so that two nodes are required to have at least a threshold number of keys in common before they can derive a key.
\end{description}
These baseline KPSs provide suitable motivation for several observations concerning the building of KPSs for WSNs:
\begin{enumerate}
\item \textit{Optimal connectivity is not necessary}. Optimal connectivity is a nice feature, but unnecessary in a KPS for a WSN. It is certainly not needed in fully controlled WSNs. However, even in uncontrolled WSNs, since only a minority of sensor nodes will be within communication range of one another, the ``costs'' of optimal connectivity might not be worth paying.
\item \textit{Deterministic schemes have some advantages}. The obvious advantage of deterministic KPSs is that we can generally make definitive statements about their properties, which aids analysis. The example of Blom's KPS also illustrates that in deterministic schemes it may be possible to have very efficient shared key discovery. Lee and Stinson \cite{lee:practical} also point out that deterministic schemes tend to involve fewer expensive pseudorandom computations during the key predistribution stage. In \cite{parkblake:ICCCN07} it was argued that in certain cases probabilistic solutions tend to converge to deterministic schemes, thus studying the latter provides valuable insight.
\item \textit{Flexibility is attractive}. An attractive feature of the random KPS is that it is highly configurable with respect to the competing requirements. Blom's KPS, for example, allows only minor tradeoffs to be made between storage and resilience.
\item \textit{Compromise is desirable}. It is unlikely that a WSN application will want the extreme tradeoffs seen in the case of the single and complete pairwise KPS. Even Blom's KPS is probably not enough of a compromise, with low storage coming at the cost of low resilience and computation requirements. Thus even if a KPS cannot offer flexibility, it is desirable that it offers a ``reasonable'' compromise between the competing requirements of Sect.~\ref{sec:reqs}.
\end{enumerate}
There have been a large number of proposals for KPSs for WSNs. These tend to either be variants of the random KPS, deterministic KPSs, proposals for combining schemes or KPSs with special properties. There are also several surveys \cite{camtepe:survey05,keithframework,newsurvey}, each of which takes a slightly different approach. We will now focus primarily on proposals that utilise combinatorial designs.

\section{Direct Application of Designs}\label{sec:direct}

Combinatorial designs are very natural objects to consider as candidate key rings for KPSs. They have the advantages of being deterministic and having rich and well understood structure. Indeed, they have been associated with the building of KPSs long before the emergence of WSNs \cite{MitchPipe87}. In this section we consider the direct application of $(v,b,r,k)$-designs as key rings for a KPS.

\subsection{Two Interesting Classes}

The basic definition of a $(v,b,r,k)$-design is too general to guarantee any interesting connectivity or resilience properties of the resulting KPSs. We first identify two potentially interesting classes of designs.

\subsubsection{Prioritising local connectivity:} Our first class of designs are explicitly constructed for their local connectivity properties. It was shown in \cite{lee:IEEE} that any block in a $(v,b,r,k)$-design meets (has a non-empty intersection) with at most $k(r-1)$ other blocks. Further, every block meets $k(r-1)$ blocks precisely when the design has the property that any two blocks meet in at most one point, in which case the design is known as a $(v,b,r,k)$-\textit{configuration}. These $(v,b,r,k)$-configurations are of interest since if they are used as key rings, the KPSs based on them have optimal local connectivity \cite{lee:IEEE}. Knowing that a design is a $(v,b,r,k)$-configuration does not, unfortunately, offer any immediate guarantees about its resilience.

\subsubsection{Prioritising resilience:} A class of designs with built-in resilience properties are \textit{key distribution patterns}. These were first proposed in \cite{MitchPipe87,MitchPipe88}, although we present a slightly more general definition here.
\begin{definition}\label{kdpdef}
A $w$-\emph{key distribution pattern} (KDP) is a
set system $({\I},{\B})$ with $|{\B} |=n$ such that for any pair $B_i,B_j\in\B$ with $B_i\cap B_j\not=\emptyset$ and any $\{B_{l_1},\ldots , B_{l_w}\}\subseteq {\B} \setminus \{B_i, B_j\}$, we have:
 $$B_i \cap B_j \not\subseteq
(B_{l_1} \cup \cdots \cup B_{l_w}).$$
\end{definition}
A $w$-KDP can be used as a key ring for a KPS and offers optimal resilience if no more than $w$ nodes are compromised. A $w$-KDP is only a design if it is also uniform and regular, which many known examples of KDPs are. However, Definition~\ref{kdpdef} does not provide any guarantees of connectivity.

\subsection{Fully Connected Designs}

An obvious class of designs to consider are those that offer full connectivity, which happens if every pair of blocks meet in at least one point.

\subsubsection{Fully connected configurations:} It is shown in \cite{lee:IEEE} that a $(v,b,r,k)$ configuration is fully connected precisely when it is the dual of a $2-(b,v,k,r,1)$-design. It is further shown in \cite{lee:IEEE} that when this happens, $b\leq k(k-1)+1$. Since $b$ represents the number of nodes in a WSN, and this is likely to be large, it is clear that this latter bound is one that we would like to meet, if at all possible. Fortunately there is an infinite class of configurations with this property, namely the projective planes, which are $2-(q^2+q+1,q^2+q+1,q+1,q+1,1)$-designs.

This means that if we wish to directly implement a configuration as a key ring in order to obtain a fully connected KPS with other desirable properties then there is really only one candidate family worth considering, the projective planes. Not only do they have optimal local connectivity, but amongst other advantages they have efficient shared key discovery \cite{parkblake:ICCCN07}. They were first proposed as key rings by \cite{camtepe:esorics04}. However the significant ``catch'' with using a projective plane is the restriction on the number of nodes relative to the size of the node allocation. This means that facilitating a very large number of nodes comes at the unattractive cost of relatively large key storage for each node (in this case each node allocation contains $k$ identifers, where $k$ is approximately the square root of the maximum number nodes).

\subsubsection{Fully connected KDPs:} The original concept of a KDP, as proposed in \cite{MitchPipe87,MitchPipe88}, was for fully connected KDPs. In this case a KDP, by definition, has every pair of blocks meeting in at least one point. In \cite{Stin04} structures of this type are known as $(2,w)$-KDPs. Several constructions of uniform and regular $(2,w)$-KDPs are known. In \cite{Stin97,MitchPipe88} it is shown that a 3-$(v,k,\lambda)$
-design with $w<(v-2)/(k-2)$ is the dual of a $(2,w)$-KDP. In \cite{MitchPipe88} it is shown that every $t-(v,b,r,k,\lambda)$-design is a $(2,t-2)$-KDP and every symmetric $2-(v,k,2)$-design (\textit{biplane}) is a $(2,1)$-KDP. We also observe that the complete pairwise KPS is a uniform, regular $(2,n-2)$-KDP (as well as being a configuration).

\subsubsection{Dual designs:}  By definition, the dual of a $2-(v,b,r,k,\lambda)$-design is fully connected, since every pair of blocks meet in $\lambda$ points. A special subclass are the symmetric designs, examples of which are the projective planes and the biplanes. In \cite{rujroy:ispa07} symmetric \textit{partially balanced} designs were proposed as key rings, however they share the problems of projective planes in being highly constrained in terms of the number of nodes they can support.

\subsubsection{Comment:} In general, fully connected designs are unsuitable for direct application as KPSs for WSNs. Full connectivity places too many constraints on the parameters. The main resulting problems are:
\begin{itemize}
\item \textit{Lack of flexibility}: While there are a number of constructions, they leave little room for flexibility of tradeoff between the important parameters.
\item \textit{Restrictions on number of nodes}: The tradeoff between number of nodes and storage tends to be unsatisfactory, with reasonable storage limitations leading to too tight a restriction on the maximum number of nodes.
\item \textit{Too much to the extreme}: Full connectivity provides better connectivity than we typically need for a WSN. The cost in terms of storage and resiliency is too high to be worth paying for most WSNs.
\end{itemize}
Nonetheless, direct application of designs in this way provides more baseline KPSs with special properties for comparison, as well as being potentially useful components in more complex constructions.

\subsection{Designs Without Full Connectivity}\label{sec:deswocon}

Given that full connectivity is not necessary for a KPS for a WSN, it is worth considering direct application of designs that do not have full connectivity. In such designs there will be blocks that do not intersect. This means that there will be pairs of nodes who do not share a key in the resulting KPS. In the first instance it seems wise to consider configurations, since these at least offer optimal local connectivity.

\subsubsection{Generalised Quadrangles:} In \cite{camtepe:esorics04} the use of \textit{generalised quadrangles} as WSN key rings was considered. A GQ$(s,t)$ is a $(v,b,t+1,s+1)$ -design, where $v=(s+1)(st+1)$, $b=(t+1)(st+1)$, two points lie on at most one block, two blocks meet in at most one point, and a further property that outlaws the occurrence of ``triangles'' holds.  A GQ$(s,t)$ is thus a configuration and hence offers optimal local connectivity. In \cite{camtepe:esorics04} several GQ$(s,t)$s were shown to enable KPSs with good resilience compared to the random KPS.

\subsubsection{Common Intersection Designs:} The idea behind the use of GQ$(s,t)$'s as key rings was generalised in \cite{lee:IEEE}:
\begin{definition}\label{ciddef}
Let $(\I,\B )$ be a $(v,b,r,k)$-configuration. We say that $(\I,\B
)$ is a $(v,b,r,k,\mu)$-\emph{common intersection design} (CID) if
for any distinct pair of blocks $B_i,B_j\in {\B}$ we have:
$|\{B_k\in{\B}\,:\,B_i\cap B_k\not=\emptyset \mbox{ and }B_j\cap
B_k\not= \emptyset\}|\geq \mu.$
\end{definition}
Thus any key ring based on a CID provides the guarantee that if two nodes do not share a key, there will be at least $\mu$ nodes who could act as intermediaries in a secure two-hop path between the original nodes. From a connectivity perspective it is desirable for $\mu$ to
be as large as possible since this increases the chance that one of these intermediary nodes is within communication range. Several upper bounds on $\mu$ were established in
\cite{lee:cid} and optimal CIDs were constructed using
group-divisible designs, strongly-regular graphs and generalized
quadrangles.

\subsubsection{Transversal Designs:} A useful class of CIDs is provided by the transversal designs, since a TD$(k,n)$ is a $(kn,n^2,n,k,k^2-k)$-CID. In \cite{lee:IEEE} a particularly useful construction of TD$(k,n)$s that exist for any prime $k\leq n$ was used to construct CIDs. The resulting key rings, termed \textit{linear schemes} in \cite{lee:practical} have several interesting properties:
\begin{itemize}
\item The values of $k$ and $n$ can be varied to produce key rings with a range of compromises between the storage $k$, maximum number of nodes $n^2$, local connectivity $k\/ (n+1)$ and resilience.
\item Local connectivity and resilience can be computed using formulae that were derived in \cite{lee:practical}.
\item They have a very efficient shared-key discovery phase, which involves two nodes exchanging identifiers and making a simple computation.
\end{itemize}

\subsubsection{Generalised Transversal Designs:} An example of a class of designs that do not offer full connectivity and are not configurations are the generalised transversal designs TD$(t,k,n)$. For example, a block in a TD$(3,k,n)$ intersects other blocks in either 0, 1 or 2 points, hence is not a configuration. In \cite{lee:practical} TD$(3,k,n)$s were used to construct key rings based on a requirement that a pair of nodes shares two keys before they can derive a session key. The performance of these so-called \textit{quadratic schemes} was analysed in \cite{lee:practical} and shown to offer some interesting tradeoffs. For example, they offered better resilience than linear schemes for low levels of compromised nodes, while providing similar levels of local connectivity.

\subsubsection{Trivial KDPs:} Let $\G$ be a connected graph on $n$ vertices with no loops or multiple edges. Associate a vertex with node $U_j$, assign a unique key identifier $x_i$ to each edge, and define node allocation $B_j$ to be the the set of edges (key identifiers) adjacent to $U_j$. The result is an $(n-2)$-KDP, which offers the maximum possible resilience. This is an example of a \textit{trivial inclusion}-KDP \cite{Martin:BCC}. The advantage of designing KPSs in this way is that $\G$ can be analysed for connectivity and path-length properties. Trivial KDPs are $(v,b,2,k)$ -designs when $\G$ is regular of degree $k$.

In \cite{lee:sac04} it was pointed out that one class arise from strongly regular graphs, since these graphs offer a guaranteed number of possible two-hop paths between any disconnected nodes. The cost associated with a trivial KDP is that in order to get good levels of connectivity the graph typically needs to be ``dense'' with edges, which means that the storage for each user tends to be on the high side. The \textit{IOS} KPSs in \cite{lee:sac04} employ a trick for reducing this storage which works if $\G$ is a connected regular graph whose vertices have even degree. This comes at a small computational cost, as well as the security cost of relying on a hash function.

\subsubsection{Comment:} Designs without full connectivity are certainly more promising for designing KPSs for WSNs. The main advantage over fully connected designs is increased room for flexibility. The local connectivity levels can be traded off against other parameters, particularly resilience. This increased relaxation of parameters also tends to facilitate an increase in the maximum number of nodes that can be supported given a particular storage constraint. Nonetheless, a number of problems remain:
\begin{itemize}
\item \textit{Lack of flexibility}: While flexibility is generally better than for fully connected designs, it is still severely constrained by the combinatorial requirements.
\item \textit{Restrictions on number of nodes}: Despite an improvement, there is still a limit to the number of nodes that can be supported, again due to the combinatorial constraints.
\end{itemize}
Direct application of designs without full connectivity thus provides another interesting collection of KPSs. It should be noted however that in comparison to fully connected designs, these designs have not been so much studied and so useful constructions may have not yet been discovered. While their direct applicability is limited, they again provide excellent components for building more complex KPSs.

\subsection{On Direct Application of Designs}\label{sec:comment}

The rich mathematical structure of combinatorial designs makes them suitable for building KPSs with particular properties. However most interesting classes of design probably offer too much structure. Some designs offer ``all or nothing'' guarantees of properties, when a more gradual curve would be preferable. An example of this is $w$-KDPs (for small $w$), whose resilience guarantees are no longer offered when more than $w$ nodes are compromised. Designs are also uniform and regular by definition, although there is no strict need for these properties in a KPS. The main problem however is that straight application of designs tends not to provide enough flexibility to generate a wide range of KPSs suitable for different application requirements.

\section{Designs as a Building Block}\label{sec:building}

Although combinatorial designs are not always suitable for direct application as KPSs for WSNs, they are very natural objects to use as components in the construction of a KPS. The resulting KPSs can hopefully be made more flexible, while still inheriting the advantages of designs that were outlined at the start of Sect.~\ref{sec:direct}. Another way of looking at this is to start with a KPS based on direct application of a design and consider in what ways we could transform the original scheme in order to ``get more for our money''. We now consider a number of different techniques.

\subsection{Splitting a KPS}\label{sec:split}

One way of modifying a KPS is to \textit{split} nodes, by associating each node in the original KPS with a set of nodes in a new KPS. The new scheme essentially consists of $l$ versions of the original KPS. The main gain here is that this allows an $l$-fold increase in the number of possible nodes in the network compared to the original scheme.

The simplest technique is to essentially create $l$ ``mirror copies'' of the original KPS, where each split node is assigned the same node allocation of keys as its parent node. For general applications, this might seem a strange thing to do since there will now be $l$ nodes with exactly the same keys. However, for many applications of WSNs, particularly those where the main required security service is confidentiality with respect to non-members of the network, this may well be quite acceptable.

The other extreme is to associate each version of the original KPS with a disjoint set of key identifiers. This will result in a significant reduction in the connectivity, since only nodes associated with a particular version will share keys. Partially overlapping sets of key identifiers will allow tradeoff between these two extremes.

The \textit{Multiple IOS} KPSs in \cite{lee:sac04} used this idea of splitting to increase the maximum network size of the IOS KPS (see Sect.~\ref{sec:deswocon}). Since they are slightly different from our standard notion of a key rings (as defined in Sect.~\ref{sec:stages}), the cost of splitting is different (in this case it is a loss of resilience).

\subsection{Extending a KPS}\label{sec:extend}


As noted in Sect.~\ref{sec:direct}, one of the main problems with straight application of a combinatorial design as a KPS is the restriction on the maximum number of nodes. One technique for overcoming this is to generate a KPS based on a combinatorial design and then extend it by appending additional node allocations that are not part of the original scheme.

This technique was used in \cite{camtepe:esorics04} to extend KPSs based on projective planes and generalized quadrangles. In order to enforce a degree of separation between the appended node allocations and the originals, the new node allocations were selected as random subsets of blocks of the \textit{complementary design} (whose blocks are the complements of the blocks of the original design). The resulting KPSs were analysed in \cite{camtepe:esorics04} and shown to have better connectivity than a random KPS, while allowing a greater number of nodes and increased resilience in comparison to the underlying design-based KPSs.

Another possibility is to combine the node allocations of two different KPSs. This approach was taken in \cite{rujroy:ispa07}, where two KPSs arising from direct application of two partially balanced designs were combined. The resulting KPS remained fully connected, while the resiliency of the new scheme was slightly poorer than those of the original KPSs.

\subsection{Packing a KPS}\label{sec:pack}


Another option is to increase the size of node allocations by adding key identifiers (\textit{packing}). By packing the key identifiers more densely, we can expect better connectivity properties at an expected cost to resilience.

We saw in Sect.~\ref{sec:deswocon} that some combinatorial designs without full connectivity have attractive properties for adoption as key rings. However, such designs often have low inherent connectivity. In \cite{chakra:journal06} two packing strategies were tested in an attempt to increase the connectivity of the linear KPSs based on transversal designs. Both strategies involved merging KPS node allocations. The first strategy was random, whereas the second was deterministic. The results indeed indicate a small increase in connectivity at a small cost to resilience.

\subsection{Breaking up a KPS}\label{sec:break}

An alternative to making a KPS ``bigger'' through extending or packing is to break it up in various ways. Two initial suggestions for creating potentially interesting tradeoffs are:
\begin{itemize}
\item \textit{Contracting a KPS}: By removing key identifiers, either throughout the KPS or just on certain nodes, storage could be reduced at a cost to connectivity.
\item \textit{Block splitting}: By splitting node allocations (for example creating two smaller node allocations from each original node allocation by dividing it in two) the maximum network size could be increased and storage reduced, again at a cost to connectivity.
\end{itemize}
To our knowledge, the full benefits of these strategies as techniques for building KPSs with interesting properties have not yet been fully explored.

\subsection{Modifying a KPS}\label{sec:modify}


Designs can be used to make structural modifications to an existing KPS. An interesting example of this is the \textit{Modified Blom KPS} \cite{lee:sac04}. The Blom KPS, defined in Sect.~\ref{sec:baseline}, is a fully connected KPS based on a symmetric polynomial of degree $w$. In the Modified Blom KPS, we first define a complete bipartite graph on the set of nodes, which splits the nodes into two classes ${\U}_1$ and ${\U}_2$. We now establish a ``Blom KPS'' using an asymmetric polynomial (see \cite{lee:sac04} for details), which results in only pairs of nodes from distinct classes directly being able to establish a key. Nodes from the same class are required to establish a two-hop path via a node in the other class. This loss of connectivity comes at a gain in resilience, since an attacker now needs to compromise $w$ nodes from one of the classes before the KPS is completely broken.

\subsection{Joining KPSs}\label{sec:join}

A more sophisticated use of KPSs as building blocks is to join many copies of a KPS together. Of course, we need a ``rule'' to determine how the integration is done. A natural source of such a rule is another KPS, perhaps with quite different properties.
The intention is that the resulting KPS will mix the inherent properties of the component schemes.

\subsubsection{Product KPS:} In such schemes an \textit{inner} KPS and an \textit{outer} set system $({\I},{\B}^{\rm out})$ are integrated in the following way:
\begin{enumerate}
\item The outer set system provides the core structure. Each node $U_j$ is associated with the block $B_j^{\rm out}$.
\item Each key identifier $x_i$ in the set system defines a subset of nodes $N_i=\{U_j\,:\,x_i\in B_j^{\rm out}\}$. An \textit{inner} KPS is then defined on the nodes $N_i$. In this KPS, each node $U_j\in N_i$ receives the node allocation $B_j^{{\rm in}-i}$.
\item Only the node allocations of the inner KPSs are used in the final KPS. Hence each node $U_j$ receives the final node allocation $$B_j=\cup_{x_i\in B_j^{\rm out}}B_j^{{\rm in}-i}.$$
\end{enumerate}
Hence each node in the product KPS receives a final node allocation that consists of several inner KPS node allocations, one for each key identifier in the block associated with the node in the outer set system.

Note that while we have defined the Product KPS in terms of inner KPSs based on a key ring, there is no reason why other KPSs cannot be used. Indeed the low storage of the Blom KPS, which derives its key ring rather than storing it explicitly is an attractive candidate for the inner KPS, as will see in some of the following instantiations of this generic scheme.

\subsubsection{Wei-Wu Product schemes:} In \cite{wei:sac04} a general analysis of the Product KPS was conducted. It was shown that if the block size is fixed then the best resilience can be obtained if the outer set system is a design. Several constructions were proposed that used Blom KPSs as the inner KPS and used designs based on \textit{difference sets} as the outer set system.

\subsubsection{Multiple Space Blom scheme:} In \cite{lee:practical} a product KPS was proposed where the outer set system is a linear KPS (see Sect.~\ref{sec:deswocon}) and the inner KPSs are Blom KPSs. The resulting KPS was shown to have a different resilience curve compared to a linear scheme (better resilience for small numbers of compromised nodes) at the cost of some computation in order to establish keys. The efficient shared-key discovery property of both components is preserved.

\subsubsection{Multiple Space Modified Blom scheme:} In \cite{lee:sac04} a product KPS was proposed where the outer set system consists of a trivial KDP based on a strongly regular graph that has been split into $l$ identical copies (as in Sect.~\ref{sec:split}) and the inner KPSs are Modified Blom KPSs. The Modified Blom KPSs were applied using the natural partition defined by the two classes of split nodes adjacent to each edge in the strongly regular graph. The resulting scheme was shown to have a different resilience curve compared to deploying a Modified Blom KPS across all nodes.

\subsubsection{Park-Blake schemes:} In \cite{parkblake:ICCCN07} complete subgraphs of two strongly regular graphs (the \textit{triangular graph} and \textit{lattice graph}) were used to define outer set systems. It was shown that if projective planes are used as the inner KPSs then the new schemes allow a greatly increased network size while still gaining from the efficient shared key discovery of the projective plane. Clearly these constructions also lend themselves to use of Blom schemes as the inner KPSs.

\subsubsection{Scope for Joining KPSs:} Joining KPSs seems to be an interesting way of generating KPSs with different parameter tradeoffs. There seems to be plenty of scope for further exploring effective ways of combining KPSs, as most of the existing work has focussed on instantiations of the Product KPS and employed Blom's KPS as the inner KPSs.

\subsection{The Pros and Cons of Combinatorial Engineering}

In Sect.~\ref{sec:direct} we saw that direct application of combinatorial designs is generally too restrictive to produce KPSs that are suitable for WSNs. In this section we have discussed a number of different techniques for using KPSs based on designs as building blocks. There is no reason why these techniques could not be used to combine deterministic KPSs based on designs with probabilistic random KPSs.

With the exception of layered KPSs, this ``combinatorial engineering'' is not a normal study area for pure mathematicians and hence little theory on the subject exists. Indeed, for many of the techniques, the underlying combinatorial structure is sufficiently destroyed that the resulting properties can only be determined by simulations.

It might be felt that combinatorial engineering is self-defeating in that many of the advantages of using combinatorial designs may be lost, especially if they are combined with probabilistic KPSs. However it would seem that some combinatorics can be better than no combinatorics, since the properties of the underlying design-based KPS in most cases still provides some structural guarantees. It is also important to keep in mind the observations made in Sect.~\ref{sec:baseline} and \ref{sec:comment}, which indicate that KPSs for WSNs are not by definition classical combinatorial objects and thus lend themselves to this type of manipulation.

\section{Designs for Special Networking Environments}\label{sec:special}

In this section we examine the application of combinatorial designs to KPSs that do not fully conform to the application environment of uncontrolled homogeneous nodes that we have discussed thus far.

\subsection{Partially Controlled KPSs}\label{sec:partial}

The KPSs that we have discussed thus far are all uncontrolled (see Sect.~\ref{sec:wsn}) with respect to their final location. If we are able to have partial control over the location of nodes then this knowledge can be very useful in building a suitable KPS.

An example of partial control occurs in networks in which
nodes are deployed in groups in such a way that nodes from a group are deployed closer
together on average than nodes from different groups. This \textit{group deployment} might arise, for example, if nodes are deployed in batches from an aeroplane. It would be reasonable to expect nodes from one group to then be physically located closer to one another than nodes from different groups. As a result, keys can be predistributed more efficiently if this knowledge is taken into account.

A possible paradigm is to assign node allocations to each group using a KPS defined only on that group. This KPS could be more ``relaxed'' with respect to connectivity than an uncontrolled KPS. We then need to build in some means for nodes from different groups to establish common keys. However it is also important to avoid communication bottlenecks, or the risk that an entire group could become disconnected from the rest of the network, so it is desirable to ensure that the probability of nodes from different groups being able to communicate securely is similar to that of nodes from within a group. This property was referred to as {\em balanced local connectivity} in \cite{martinpatersonstinson:groups}.

The inherent structure required for group deployment of a KPS that has balanced local connectivity lends itself naturally to use of a combinatorial structure. In \cite{martinpatersonstinson:groups} such a KPS was proposed that utilises the structure of a resolvable transversal design TD$(k,m)$. The $m$ parallel classes $P_1,P_2,\ldots ,P_m$ of blocks of this design are further partitioned into $\mu$ sets of parallel classes $S_1,S_2,\ldots ,S_{\mu}$, each containing $m /\mu$ blocks. Nodes in group $G_i$ are associated with the blocks in parallel classes contained in $S_i$. The proposed KPS is based on the Multiple Space Blom scheme of Sect.~\ref{sec:join}, with the outer KPS being a based on the resolvable TD$(k,m)$. However in this KPS there are two inner KPSs:
\begin{enumerate}
\item As in the Multiple Space Blom scheme, for each key identifier $x_i$ in the outer KPS, a Blom KPS is defined on the nodes $N_i=\{U_j\,:\,x_i\in B_j^{\rm out}\}$.
\item A further Blom KPS is defined on the set of nodes $M_i=\{U_j\,:\,B_j^{\rm out}\in P_i\}$.
\end{enumerate}
The analysis of the resulting KPS in \cite{martinpatersonstinson:groups} shows that it offers good balanced connectivity, while providing a flexible set of configurable parameters that allow connectivity and resilience to be traded off against storage costs. The KPS also inherits the efficient shared key discovery of the underlying outer KPS based on the transversal design.

\subsection{Fully Controlled KPSs}\label{sec:full}

It is of significant advantage for key predistribution to know the precise deployment location of nodes, since this is even more useful than partial location information, as discussed above. It might seem that in such cases of fully controlled networks it suffices to issue a node with keys for each of its neighbours, since these are known in advance. However in dense networks this is an inefficient technique and there are much better options.

If nodes are deployed in a highly structured physical formation then it again becomes natural to look to combinatorial mathematics for building KPSs. The case of KPSs for WSNs arranged in square and hexagonal grids has been investigated in \cite{gbkps} and \cite{khop}. Efficient KPSs were constructed using a special type of combinatorial structure called a \textit{distinct difference configuration}. While these are not combinatorial designs, the node assignments that they generate can be viewed as a type of infinite combinatorial design. Nonetheless, this example serves a warning that there are special networking environments where the ``right'' combinatorial structure for building efficient deterministic KPSs is not necessarily based on a conventional combinatorial design.

\subsection{KPSs for Heterogeneous Networks}\label{sec:hetero}

Although we restricted our previous discussion to homogeneous networks (see Sect.~\ref{sec:wsn}) it is worth making some observations about \textit{heterogeneous networks}, where not all the nodes have the same capabilities. The most interesting class of heterogeneous network is probably \textit{hierarchical networks}, where the nodes are partitioned into an ordered hierarchy, with nodes at a given level being more powerful than nodes at lower levels. The most common scenario is a simple \textit{two-level hierarchy}. We now make a few observations about the applicability of combinatorial designs to building KPSs for heterogeneous networks, and in particular two-level hierarchies.

\subsubsection{Simple two-level hierarchies:} It is worth observing that many of the manipulations of KPSs discussed in Sect.~\ref{sec:building} can result in KPSs that are suitable for simple two-level hierarchies, since the resulting node allocations have different sizes. For example:
\begin{itemize}
\item A KPS could be partially packed (see Sect.~\ref{sec:pack}) using a strategy that results in node allocations of two different sizes, the original and the packed. Nodes with packed node allocations will require greater storage capability. Further, as they hold more keys it is reasonable to expect them to be more likely to be involved in communication (both directly and as an intermediary). The resulting KPS is suited to applications where there is a two-level hierarchy of sensors where there is a fairly small difference in capability.
\item Similarly, if a KPS is extended (see Sect.~\ref{sec:extend}) by adding node allocations of a different size to the original, then the resulting KPS will have similar properties to the previous case.  For example, the extension to the projective plane discussed in \cite{camtepe:esorics04} could involve choosing larger subsets of the complementary design, hence creating two classes of node allocations.
\end{itemize}

\subsubsection{Two-level hierarchies with a backbone:} A more sophisticated class of two-level hierarchies are formed by networks where the top level of nodes form a fully connected \textit{backbone}. Low level nodes are organised into \textit{subnetworks} (sometime called \textit{clusters}) which ``hang off'' this backbone and are each associated with a unique high level node. Two low level nodes from the same subnetwork can try to communicate directly. On the other hand, two low level nodes from different subnetworks have to communicate via their high level node representatives. The top level nodes thus need to be significantly more powerful than low level nodes, since they are used as communication intermediaries. It is thus also reasonable to assume that they have significantly increased storage capability.

There are many different possible approaches to designing a two-level hierarchical KPS with a backbone. The fully connected backbone could be realised, for example, by any of the KPSs discussed in Sect.~\ref{sec:full}. The subnetworks could be instantiated by any KPS (based on a design, or otherwise). There has been a significant amount of general research on key management in two-level hierarchical networks but, with the exception of a simple framework proposed in \cite{chakraseberry}, there has been very little analysis of how to build deterministic two-level KPSs. There seems to be plenty of scope for further examining exactly how best to choose both the backbone and subnetwork KPSs in order to achieve two-level hierarchical KPSs with interesting properties. In particular, intelligent application of design-based KPSs would seem quite likely to lead to useful constructions.

\subsection{Comment}

We have seen in this section that combinatorial designs have had a role to play in building KPSs for WSNs that do not conform to the ``classical'' model of uncontrolled homogeneous nodes. They have found very natural application to group deployment of nodes, but are apparently less applicable to fully controlled deployment of nodes. What remains largely unexplored is their suitability to the design of deterministic heterogenous WSNs, and this merits further study.

\section{Concluding Remarks}

We have explored the use of combinatorial designs in building KPS for WSNs. While designs have been widely proposed for use in such schemes, to what extent are these schemes really useful? We argued that for WSNs full connectivity is not really necessary and that a key attribute of any KPS is flexibility to allow parameter tradeoffs. This tends to rule out many straight applications of designs as KPSs, although we have seen several examples of flexible families of designs, such as transversal designs, having several useful applications. However it certainly does not rule out designs either as building  blocks or components of KPSs. The unusual combinatorial engineering techniques that have seen the basic structure of a design manipulated in order to provide more flexible KPSs are certainly interesting and merit further study, although formal theoretical analysis of such techniques (as in many engineering processes) is not always possible. Combining KPSs based on designs has proved to be a very successful strategy for obtaining deterministic KPSs that trade off parameters, however again there would seem to be more work to do in fully understanding the best combination rules. Thus we would argue that combinatorial designs most definitely do have an important role to play in building KPSs for WSNs, but that their full potential is not yet fully understood.

\bibliography{iwcc}

\end{document}